\documentclass[aps,prb,twocolumn,letterpaper,showpacs]{revtex4}
\usepackage[utf8]{inputenc}
\usepackage{graphicx,amsmath}
\usepackage{amssymb}
\usepackage{amsthm}

\usepackage{xcolor}

\newcommand{\bsigma}{\mbox{\boldmath$\sigma$}}

\bibpunct{[}{]}{,}{n}{}{}

\begin{document}

  \title{Electron-hole pairing of Fermi-arc surface states in a Weyl-semimetal bilayer}
  \author{Paolo~Michetti}
  \email{paolo.michetti@tu-dresden.de}
  \affiliation{Institute of Theoretical Physics, Technische Universit\"at Dresden, 01062 Dresden, Germany}
  \author{Carsten~Timm}  
  \email{carsten.timm@tu-dresden.de}
  \affiliation{Institute of Theoretical Physics, Technische Universit\"at Dresden, 01062 Dresden, Germany}
  
  \date{December 23, 2016}
  
  \begin{abstract}
    The topological nature of Weyl semimetals (WSMs) is corroborated by the presence
    of chiral surface states, which connect the projections of the bulk Weyl points by Fermi arcs (FAs).
    We study a bilayer structure realized by introducing a thin insulating
    spacer into a bulk WSM.
    Employing a self-consistent mean-field description of the interlayer
    Coulomb interaction, we propose that this system can develop an
    interlayer electron-hole pair condensate.  
    The formation of this excitonic condensate
    leads to partial gapping of the FA dispersion.
    We obtain the dependence of the energy gap and the critical
    temperature on the model parameters, finding, in particular, a
    linear scaling of these quantities with the separation between the Weyl points
    in momentum space. 
    A detrimental role is played by the curvature of the FAs, although the 
    pairing persists for moderately small curvature.
    A signature of the condensate is the modification of the
    quantum oscillations involving the surface~FAs.
  \end{abstract}

 \maketitle


\section{Introduction}

By now, a good understanding of band-topological gapped phases has been reached. 
These include the 2D integer quantum Hall state and the 2D and 3D topological insulating (TI)
phases~\cite{XLReview2010,HasanKane,chiu}.
More recently, it has been realized that also metallic systems can have band-topological properties. 
3D Weyl semimetals represent an important example~\cite{murakami, wan, burkov, turner, WCK14, VaV14}.
They are characterized by nondegenerate, linearly dispersing (Weyl) cones in their
low-energy dispersion, where the Weyl cones come in pairs
of opposite chirality $\kappa=\pm 1$ separated in momentum space. 
It can easily be shown that the Weyl nodes act as monopoles of Berry curvature,
carrying a Berry flux $2 \pi\,\kappa$.
The stability of the WSM phase is guaranteed by the conservation of the total Berry flux.
As a consequence, a Weyl node can be gapped out only by merging it
with a Weyl node of opposite chirality.

Murakami~\cite{murakami} has shown that the WSM appears as an intermediate phase 
between the normal insulating (NI) 
and the TI phases in 3D materials with broken inversion symmetry, while the topological
transition is sharp in presence of inversion symmetry.
More generally, a WSM can always be realized starting from a 3D Dirac
system~\cite{liu2014a,liu2014b} by breaking either inversion or time-reversal symmetry.
Possible realizations have been proposed in a variety of systems
such as TI/NI heterostructures~\cite{burkov}, pyrochlore iridates~\cite{wan},
TlBiSe$_{2}$\cite{singh}, HgCr$_2$Se$_4$~\cite{xu}, TaAs~\cite{huang,weng},
non-cen\-tro\-sym\-me\-tric monophosphides~\cite{weng},
and Co-based Heusler compounds~\cite{wang2016}.
Recently, WSM phases with broken inversion symmetry have been identified
in TaAs~\cite{xu2,yang}, NbP~\cite{souma}, and MoTe$_2$~\cite{ke}.

The bulk physics of WSMs reveals interesting transport phenomena 
such as negative magnetoresistance, the chiral magnetic effect, and the
quantum anomalous Hall effect, which are related to the chiral
anomaly~\cite{zyuzin,liu2013,landsteiner}.
The surface of a WSM also exhibits interesting physics in that
it supports surface states~\cite{wan,huang,yang}, which are
closely related to the topological nature of this phase.
When the chemical potential is tuned to the Weyl nodes, the bulk Fermi surface is solely given
by pairs of Weyl points, which are connected by FAs of chiral surface states.

The study of interacting states with non-trivial topology~\cite{chiu} is currently
one of the most active research areas in condensed matter physics.
One possibility in this context is that interactions induce symmetry-breaking
phase transitions. Symmetry breaking can in principle affect either the bulk topological
material and, due to bulk-boundary correspondence, then also its surface or it can happen only
at the surface. In either case, it may lead to gapping of symmetry-protected topological
surface states.
For example, excitonic phases emerging in a bulk WSM phase have been studied in Ref.~\onlinecite{wei}.
More complex situations such as interactions between two adjacent surfaces are also
of interest. 
A number of groups~\cite{TECMoore,HaoTEC,ChoMooreMagnetic,tilahun2011,Efimkin2012}
have considered the particle-hole pairing between surface states of 3D TIs, which realize an electron-hole bilayer.

In fact, the possibility of a macroscopically coherent electron-hole state, i.e., an
exciton condensate~\cite{blatt1962,moskalenko1962,keldysh1964}, due to 
the interaction between electrons and holes 
was studied much earlier, in bilayer semiconductors structures.
The original idea \cite{lozovik1976} was to employ such structures to 
overcome detrimental interband processes~\cite{shevchenko1994} and radiative recombination,
which affect the stability of an exciton condensate in bulk crystals.
In the last twenty years, exciton condensation has been studied in bilayer quantum-well 
structures, particularly in the quantum Hall
regime~\cite{girvin,dasgupta2011,snoke2011,eisenstein}.
Recent research on exciton condensates addresses
bilayer Dirac systems such as two-layer
graphene~\cite{lozovik2008,zhang2008,min2008,kharitonov2008} 
and double quantum wells embedded in semiconductor heterostructures with strong spin-orbit
interaction~\cite{can2009,HaoChiralTEC,budich2014}.

It is then interesting to see whether exciton condensation is possible for
the chiral Fermi arc states (FASs) at the surface of a WSM.
In this work, we answer this question in the affirmative. 
We study the instability towards electron-hole pairing of FASs for the case
where two parallel surfaces are formed by inserting an insulating 
spacer of thickness $t$ into a WSM crystal.
In Sec.~\ref{sec:WSM}, we present a minimal WSM model with straight FAs and solve it for a slab geometry, 
obtaining the energy-dispersion curves and the wave functions of its FASs.
In Sec.~\ref{sec:MF}, a mean-field (MF) treatment of the electron-hole
pairing in bilayer systems
is introduced, which we apply to the case of a WSM bilayer in Sec.~\ref{sec:eh_pairing}.
In Sec.~\ref{sec:numerics}, we present the numerical solutions of the gap equation derived
in the MF approximation and analyze the dependence of the excitonic gap and the critical
temperature on the model parameters.
We then discuss the role of chemical doping and a inter-layer potential bias.
As a step beyond the minimal WSM model of Sec.~\ref{sec:WSM}, we also address curved FAs.
We show that the phenomenon persists for moderate curvature of the FAs.

\section{Model and theory}

\subsection{Minimal model for Weyl semimetals\\ and Fermi arcs}
\label{sec:WSM}

We construct a minimal WSM model starting from
the Dirac equation for a free particle in the Weyl representation~\cite{peskin},
which can be written as
\begin{equation}
H_0 \psi = E \psi,
\label{eq:Dirac}
\end{equation}
where
\begin{equation}
H_0 = - v_F {\mathbf k} \cdot \bsigma\otimes \tau_z
  + M\, \sigma_0\otimes\tau_x .
\end{equation}
Here, $\sigma_j$ ($\tau_j$) are the Pauli matrices and $\sigma_0$ ($\tau_0$)
is the unit matrix referring to the Weyl cone (particle-hole sector).
We set $\hbar=1$.
Additionally, we introduce the time-reversal-symmetry-breaking
term $H'= v_F\, K_0\, \sigma_x\otimes\tau_0$, 
corresponding to the coupling with a Zeeman field along $\hat{\mathbf{x}}$,
which lifts the degeneracy between right-handed (\textit{R}) and left-handed (\textit{L}) states.
For $M=0$, the Zeeman term splits the Dirac node, which is originally located
at ${\bf k} = 0$, into two Weyl nodes at $k_x=\pm K_0$ for \textit{L} and \textit{R} states, 
respectively, leading to a WSM phase. 
The WSM phase persists for finite values of $M$ due to the topological protection of Weyl
nodes as long as the the Zeeman term is dominant~\cite{murakami}.

Now, we solve the eigenvalue problem of $H=H_0+H'$ for a planar interface at $z=0$,
separating a vacuum region with $K_0=0$ and $M\rightarrow\infty$
from a WSM domain with $K_0\ne 0$ and, for the sake of simplicity,
$M=0$.
In the WSM phase, $H$ decouples into two Weyl equations for \textit{L} and \textit{R}
fermions,
\begin{equation}
H_{L,R} = - v_F\, \left[(k_x \mp K_0) \sigma_x + k_y\sigma_y+k_z\sigma_z\right],
\label{eq:H_LR}
\end{equation}
where the sign $-$ ($+$) stands for the \textit{L} (\textit{R}) sector. 
$k_x$ and $k_y$ are still good quantum numbers, while values of $k_z$
compatible with a fixed energy $E$ are obtained by 
the dispersion relation (secular equation) of Eq.~(\ref{eq:H_LR}).
We search for the FASs in the energy range
$E^2 < v_F^2\, [(k_x \pm K_0)^2+k_y^2]$, where $k_z$ assumes imaginary values.
Matching the evanescent modes compatible with the energy $E$ at the interface
between the WSM and the vacuum, 
we obtain, in the wave-vector range $k_x\in(-K_0,K_0)$, FAS solutions with
\emph{unidirectional} dispersion relation 
\begin{equation}
\varepsilon^{(\pm)}_{\bf k} = \pm v_F\,k_y,
 \label{eq:dispFASS}
\end{equation}
where $\pm 1$ identifies the possible surface normal $\pm \hat{\mathbf{z}}$
of the WSM domain and $\mathbf{k}=(k_x,k_y)$ is the wave vector parallel to the surface.
The corresponding FAS wave functions are given by   
\begin{equation}
 \Psi^{(\pm)}_{\mathbf{k}}({\bf r})
   = \frac{e^{i(k_x x+k_y y)}}{\sqrt{L_xL_y}}\, \sqrt{\frac{K_0^2-k_x^2}{2K_0}} \,
     \left(\begin{array}{@{}c@{}}
     e^{\pm(K_0-k_x)z} \\ \mp i\,e^{\pm(K_0-k_x)z} \\ \pm i\,e^{\pm(K_0+k_x)z} \\
     -e^{\pm(K_0-k_x)z}
  \end{array} \right).
  \label{eq:FA_wavef}
\end{equation}

\subsection{Mean-field treatment of bilayer\\ electron-hole pairing}
\label{sec:MF}
  
In this section, we briefly summarize the MF theory of electron-hole pairing in a bilayer
structure~\cite{lozovik1976,littlewood1996}
with layers \textit{A} and \textit{B} and one energy band per layer. 
Recall that the bands of the WSM are non-degenerate with spin locked to momentum.
We express the inter-layer Coulomb interaction as
\begin{equation}
\hat H_C = - \sum_\mathbf{Q}\, \sum_{\mathbf{k} \ne \tilde{\mathbf{k}}}
  V_{\mathbf{k},\tilde{\mathbf{k}}}^\mathbf{Q}
  \hat P_{\mathbf{k},\mathbf{Q}}^\dag \hat P_{\tilde{\mathbf{k}},\mathbf{Q}},
\label{eq:H_C_pair}
\end{equation}
where we have introduced the pair operator
$\hat P_{\mathbf{k},\mathbf{Q}} = \hat b^\dag_{\mathbf{k}+\mathbf{Q}}\hat a_\mathbf{k}$,
with $\hat a$ and $\hat b$ being
electron annihilation operators for the \textit{A} and \textit{B} layers, respectively.
Within a pairing approximation, we only keep track of the interaction terms containing 
pair operators with a specific modulation vector $\mathbf{Q}$, 
most likely to realize a finite anomalous average
or electron-hole pair amplitude $\langle \hat P_{\mathbf{k},\mathbf{Q}}\rangle$.
This approximation yields a BCS-type two-band model for the inter-layer
particle-hole condensation described by the Hamiltonian 
\begin{equation}
\hat H^\mathrm{BCS}_\mathbf{Q} = \sum_\mathbf{k} \left( E_{\bf k}^{A}\, \hat a_\mathbf{k}^\dag
    \hat a_\mathbf{k} + E_{\bf k}^{B}\, \hat b_\mathbf{k}^\dag \hat b_\mathbf{k} \right)
  - \sum_{\mathbf{k}, \tilde{\mathbf{k}}} V_{\mathbf{k},\tilde{\mathbf{k}}}^{\bf Q}
    \hat P_{\mathbf{k},\mathbf{Q}}^\dag \hat P_{\tilde{\mathbf{k}},\mathbf{Q}}, 
\label{eq:HBCS}
\end{equation}
where $E_{\bf k}^{A}$ and $E_{\bf k}^{B}$ are the single-particle 
energies for the two layers, measured with respect to the chemical potential $\mu$.
The intra-layer electron-electron interaction is neglected; we effectively
assume that is does not lead to exciton formation and that the corresponding Hartree
energy is already included in $E_\mathbf{k}^A$ and $E_\mathbf{k}^B$.

The MF treatment consists of writing
$\hat P_{\mathbf{k},\mathbf{Q}} = F_{\mathbf{k},\mathbf{Q}} +
(\hat P_{\mathbf{k},\mathbf{Q}} - F_{\mathbf{k},\mathbf{Q}})$ in Eq.~(\ref{eq:HBCS})
in terms of the pair amplitude
$F_{\mathbf{k},\mathbf{Q}} \equiv \langle \hat P_{\mathbf{k},\mathbf{Q}} \rangle$
and of keeping only the first-order terms in the fluctuations.
Apart from a constant energy term, the MF Hamiltonian becomes 
\begin{align}
\hat H^\mathrm{MF}_\mathbf{Q} =& \sum_\mathbf{k} \left( E_{\bf k}^{A}\,
  \hat a_\mathbf{k}^\dag \hat a_\mathbf{k} 
  + E_{\mathbf{k} + \mathbf{Q}}^{B}\, \hat b_{\mathbf{k} + \mathbf{Q}}^\dag
    \hat b_{\mathbf{k} + \mathbf{Q}} \right) \nonumber\\ 
  &{}+ \sum_\mathbf{k} \left(\Delta_{\mathbf{k},\bf Q}^* \hat P_{\mathbf{k},\mathbf{Q}}
    + \Delta_{\mathbf{k},\bf Q} P_{\mathbf{k},\mathbf{Q}}^\dag \right), 
\label{eq:HMF}
\end{align}
where the gap parameter must be self-consistently determined
through the gap equation
\begin{equation}
\Delta_{\mathbf{k},\bf Q} = - \sum_{\tilde{\mathbf{k}}}
  V_{\mathbf{k},\tilde{\mathbf{k}}}^{\bf Q} F_{\tilde{\mathbf{k}},\mathbf{Q}}.
\end{equation}
Note that this corresponds to the condensation of excitons
of finite momentum ${\bf Q}$. 
Due to the presence of the condensate, the momentum of the single-particle states is only conserved
up to integer multiples of ${\bf Q}$.
This situation is analogous to the spontaneous breaking
of translational invariance with the emergence of a density wave,
and leads to the folding of single-particle energy dispersion into bands. 
For sufficiently large ${\bf Q}$, the mixing with higher energy bands can be 
neglected when addressing low-energy properties. 
On the other hand, for ${\bf Q}=0$, there is no folding.
We will therefore restrict ourselves to a two-band model.
The MF theory is equivalent to BCS theory for spinful electrons in a magnetic field,
as can be seen by performing the mapping $\hat c_{\mathbf{k}\uparrow}\equiv a_\mathbf{k}$,
$\hat c_{\mathbf{k}\downarrow}\equiv b_{\mathbf{k}+\mathbf{Q}}^\dag$.
The single-particle eigenenergies of $\hat H_\mathbf{Q}^\mathrm{MF}$ are
\begin{equation}
E_\mathbf{k}^{\pm} = \frac{E_{\bf k}^{A} + E_{{\bf k}
  +\bf Q}^{B}}{2} 
  \pm \bigg[\bigg(\frac{E_{\bf k}^{A} - E_{{\bf k}+\bf Q}^{B}}{2}\bigg)^{\!\!2}
  + |\Delta_{\mathbf{k},\mathbf{Q}}|^2 \bigg]^{1/2},
  \label{eq:disp_qp}
\end{equation}
and the gap equation becomes
\begin{equation}
\Delta_{\mathbf{k},\bf Q} = - \sum_{\mathbf{k}'}
  V_{\mathbf{k},\mathbf{k}'}^{\bf Q} 
    \frac{\Delta_{\mathbf{k}',\bf Q}
    \left[f{(E_{{\bf k'}}^{+})}-f{(E_{{\bf k'}}^{-})} \right]}
  {\sqrt{\left(E_{{\bf k'}}^{A} - E_{{\bf k'}+\bf Q}^{B}\right)^{\!2} 
  + 4\, |\Delta_{\mathbf{k}',\bf Q}|^2}} ,
\label{eq:Delta}
\end{equation}
where $f(E)$ is the Fermi distribution function.
We note that the order parameter of the bilayer system can be described in terms of
the layer pseudo-spin
\begin{equation}
  {\bf M}({\bf r}) \,=\, \sum_{\alpha,\beta} \bsigma_{\alpha\beta}\, \langle \hat\Psi_\alpha^\dag({\bf r}) \hat \Psi_\beta({\bf r}) \rangle,
\end{equation}
where $\bsigma$ is the vector of
Pauli matrices associated with the layer (\textit{A}, \textit{B})
degree of freedom and $\hat \Psi_1$ and $\hat \Psi_2$
are the field operators in the \textit{A} and \textit{B} layers, respectively.
For the pair amplitude $F_{\mathbf{k},\mathbf{Q}}\neq 0$, the pseudo-spin
$\mathbf{M}$ lies in the \textit{xy} plane and the phase
of $F_{\mathbf{k},\mathbf{Q}}\neq 0$ denotes the orientation of $\mathbf{M}$ within
this plane, as found for quantum Hall bilayers~\cite{moon,girvin}.
The particular case of a pair amplitude with finite ${\bf Q}\ne 0$ corresponds to the
formation of electron-hole pairs having finite total momentum.
Similar scenarios have been predicted to occur in electron-hole bilayers
which, unlike our case, have a large density
imbalance between electrons and holes~\cite{pieri,parish}.
Excitonic states with finite momentum have been invoked to explain the bulk magnetic
state of chromium~\cite{Ric70} and more recently iron pnictides~\cite{CEE08,HCW08,CvT09,BrT09}.
This scenario is also reminiscent of the Fulde-Ferrel-Larkin-Ovchinnikov (FFLO)
superconducting phase~\cite{FFLO1,FFLO2},
where translational invariance is broken by the spatial modulation of the order parameter.

\subsection{Electron-hole pairing in Fermi-arc surface states}
\label{sec:eh_pairing}

\begin{figure}[tb]
  \centering	
  \vspace{0.3cm}
  \includegraphics[width=6.5cm]{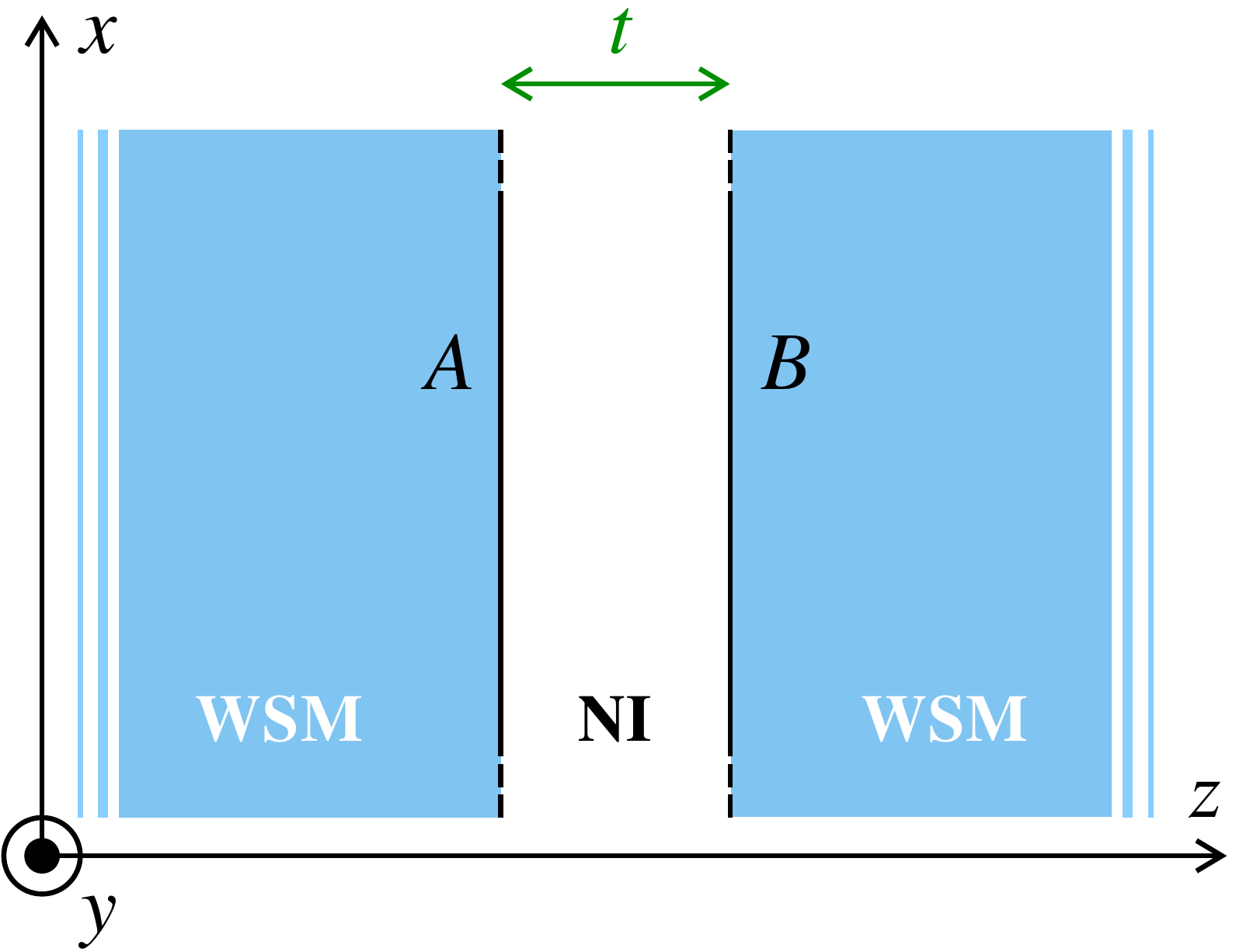}
  \caption{Illustration of a WSM bilayer, where two macroscopic slabs featuring an
  identical WSM phase are stacked in the \textit{z}-direction, separated by an
  insulating spacer of thickness $t$.}
  \label{fig:system}
\end{figure}

We now apply the MF treatment to the WSM bilayer sketched in Fig.~\ref{fig:system},
where the WSM phases are identical and described by the minimal model in Sec.~\ref{sec:WSM}.
Opposite surfaces of the two WSM slabs with normal directions
$\hat{\mathbf{n}}=\pm \hat{\mathbf{z}}$
are facing each other at a distance $t$, separated by an insulating spacer.
We take into account only surface states belonging to the FAs of the material, neglecting the 
bulk bands, which have a vanishing density of states at charge neutrality,
where the Fermi surface corresponds to two isolated Weyl points.
The FAs terminate in the projections of the Weyl points into the 2D Brillouin zone of the system.
The summations over the wave vector appearing in Eqs.~(\ref{eq:H_C_pair}) and
(\ref{eq:HBCS}) and in all other equations describing this system 
are therefore limited to states with $k_x,\tilde{k}_x,k_x+Q_x,\tilde{k}_x+Q_x\in(-K_0,K_0)$.
Here, we consider the case that there is no electrostatic potential difference 
and no difference in chemical potential between the two interfaces.
The effects of deviating from these assumptions will be discussed below.
Hence, we have $E_{\bf k}^{A}=\varepsilon_{\bf k}^{(+)}-\mu$ and $E_{\bf k}^{B}= \varepsilon_{\bf k}^{(-)}-\mu$.
The surface states have the combined symmetry of electron-hole inversion 
(charge conjugation) times mirror reflection at the center plan of the insulating
spacer. 
In addition, the system has mirror symmetries in the \textit{xz} and \textit{yz} planes.
The FAs of the WSM layers \textit{A} and \textit{B} are perfectly nested with the nesting
vector $\mathbf{Q}=0$. 
We therefore expect a uniform pairing state with $\mathbf{Q}=0$.

\begin{figure}[tb]
\centering
\includegraphics[width=7.5cm]{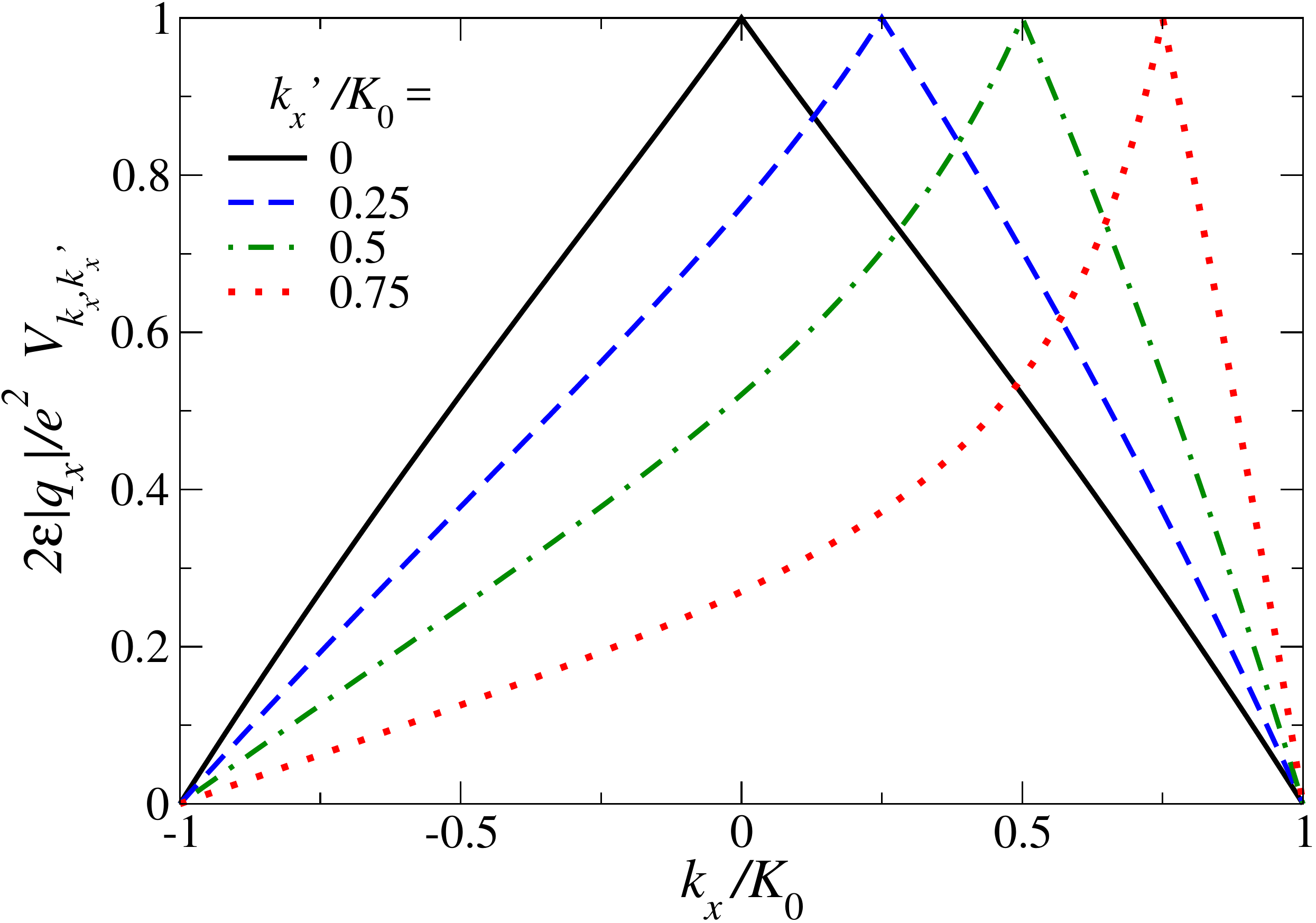}
\caption{Coulomb matrix element $V_{\mathbf{k},\mathbf{k}'}^{\mathbf{Q}=0}$ for 
$\mathbf{k}$ and $\mathbf{k}'$ on the \textit{x}-axis, keeping $k_x'$ fixed
and varying $k_x\in(-K_0,K_0)$.
The interaction becomes asymmetric and decays faster as a function of $|k_x-k_x'|$
for $k_x'$ approaching $K_0$ due to the increasing spatial extent of the FASs, which merge with the bulk states at $\pm K_0$.}
\label{fig:Vkkp}
\end{figure}

The FASs are not perfectly localized at the surface but rather decay exponentially
into the bulk, as described by Eq.~(\ref{eq:FA_wavef}). 
Therefore, matrix elements of the Coulomb potential between FASs should take into account their actual shape.
Details on the calculation are relegated to App.~\ref{app:V},
where we derive the analytical form of the Coulomb matrix element 
\begin{align}
 V_{ {\bf k} , {\bf k'} }^{Q_x=0} =& \frac{2e^2}{\epsilon_0\epsilon_r}\,
   \frac{e^{-|{\bf k'}-{\bf k}|t}}{|{\bf k'}-{\bf k}|}\,
   \frac{(K_0^2-k_x'^{2})(K_0^2-k_x^2)}{K_0^2} \nonumber\\
  & {}\times\left[\frac{|{\bf k'}-{\bf k}|+2K_0}
    {(|{\bf k'}-{\bf k}|+2K_0)^2-(k_x+k_x')^2}\right]^2,  
\label{eq:V_Q=0}
\end{align}
which turns out to be independent of $\bf Q$ as long as $Q_x=0$.
In Fig.~\ref{fig:Vkkp}, we show cuts of the matrix element of the inter-layer Coulomb
interaction $V_{\mathbf{k},\mathbf{k}'}^{\mathbf{Q}=0}$ for fixed $\mathbf{k}'$.
The matrix element depends strongly on the \textit{x} components of the
momenta and approaches zero for $k_x,k'_x\rightarrow \pm K_0$,
where the surface states become extended and merge with the bulk states.
As a consequence, the inter-layer interaction is most effective between states in the central part of the FA.

Before discussing the numerical solution of the gap equation (\ref{eq:Delta}),
let us analyze its structure assuming a constant value of $\Delta$.
We can then divide the equation by $\Delta$.
The gap equation can be cast in the scaling form (see App.~\ref{app:I} for details)
\begin{equation}
  \frac{\pi^2 \epsilon_0\epsilon_r v_F}{e^2} =
  I\left( \frac{\Delta}{v_FK_0}, \frac{k_BT}{v_FK_0}; \frac{k_{\text{cut}}}{K_0},
  \frac{1}{K_0 t} \right),                    
\end{equation}
where $I$ is a scaling function, $\epsilon_r$ the effective dielectric constant,
and $k_{\text{cut}}$ is an ultraviolet cut-off for the $k_y$ integration, which
can be thought to physically account for a band edge or for the breakdown of
the linear dispersion in Eq.~\ref{eq:dispFASS}.
Note that for the present model this cut-off is not required to cure any divergence,
unlike for the case of a 2D isotropic linear dispersion such as in graphene.
In the limit of short FAs,
\begin{equation}
K_0 \ll k_{\text{cut}} \quad\text{and}\quad K_0t\ll 1,
\label{eq:limit}
\end{equation}
the integral $I$ depends only on its first two arguments and 
we deduce that the solution $\Delta(T)$ of the gap equation 
is then given by
\begin{equation}
\Delta(T) = v_F K_0\, D\left(\frac{k_BT}{v_F K_0}\right),
\label{eq:scaling}
\end{equation}
with a $K_0$-independent function $D$.
This condition implies that both the zero-temperature gap $\Delta_0$ and the critical
temperature $T_c$ are proportional to $K_0$.

\section{Results and discussion} 
\label{sec:numerics}

\begin{figure}[tb]
\centering
\includegraphics[width=8.6cm]{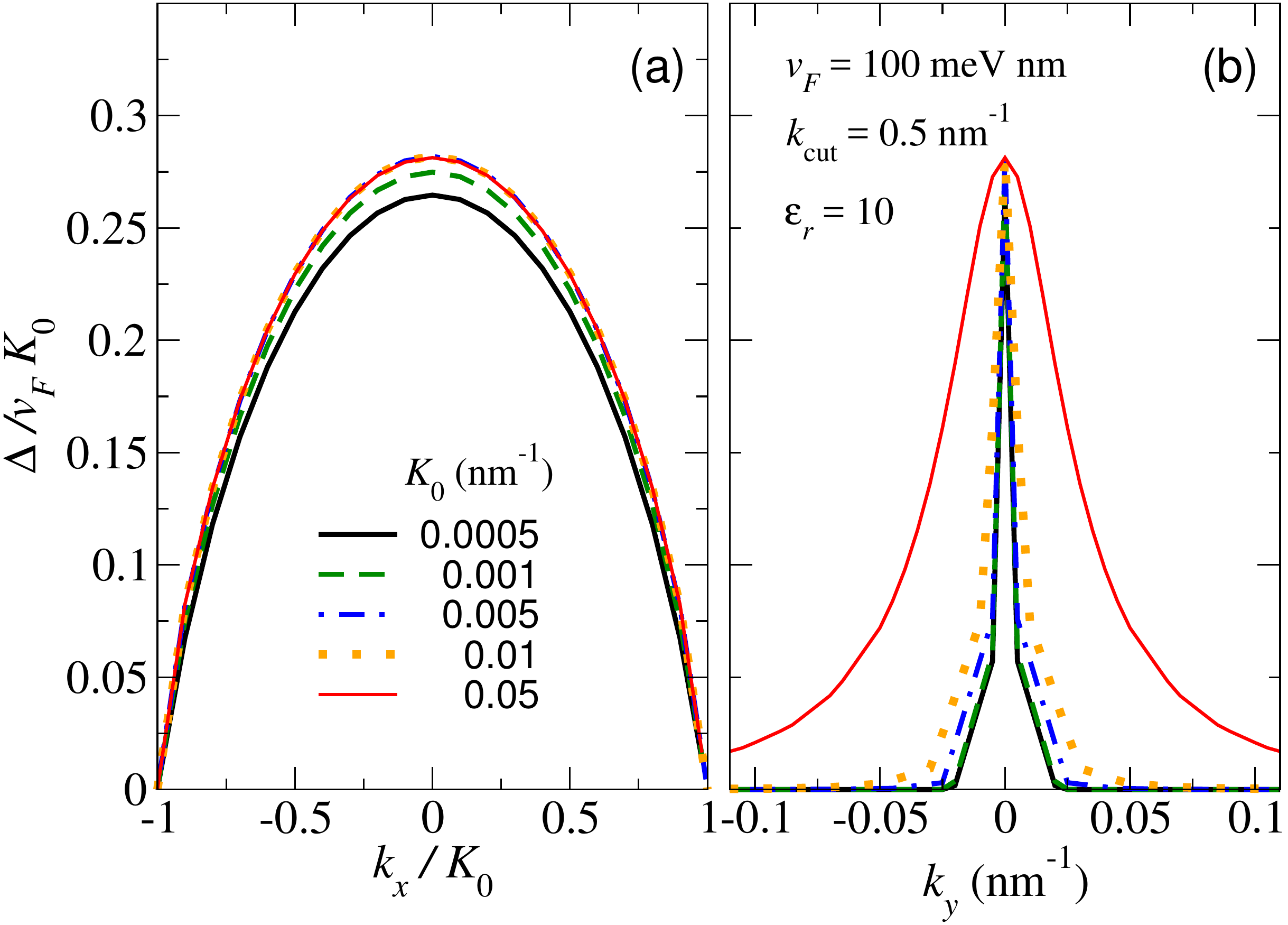}
\caption{Gap parameter (a) $\Delta_{{\bf k}=k_x \hat{\mathbf{x}}}$ and
(b) $\Delta_{{\bf k}=k_y \hat{\mathbf{y}}}$, calculated for various values of $K_0$ and $T=0$.
$\Delta_{{\bf k}=k_x \hat{\mathbf{x}}}$ scales roughly linearly with $K_0$.
$\Delta_{{\bf k}=k_y \hat{\mathbf{y}}}$ features a peak at $k_y=0$ of width on the order of $K_0$ 
and approaches zero for $|k_y|\gg K_0$.
}
\label{fig:D_kxky}
\end{figure}

\begin{figure}[tb]
\centering
\includegraphics[width=8.6cm]{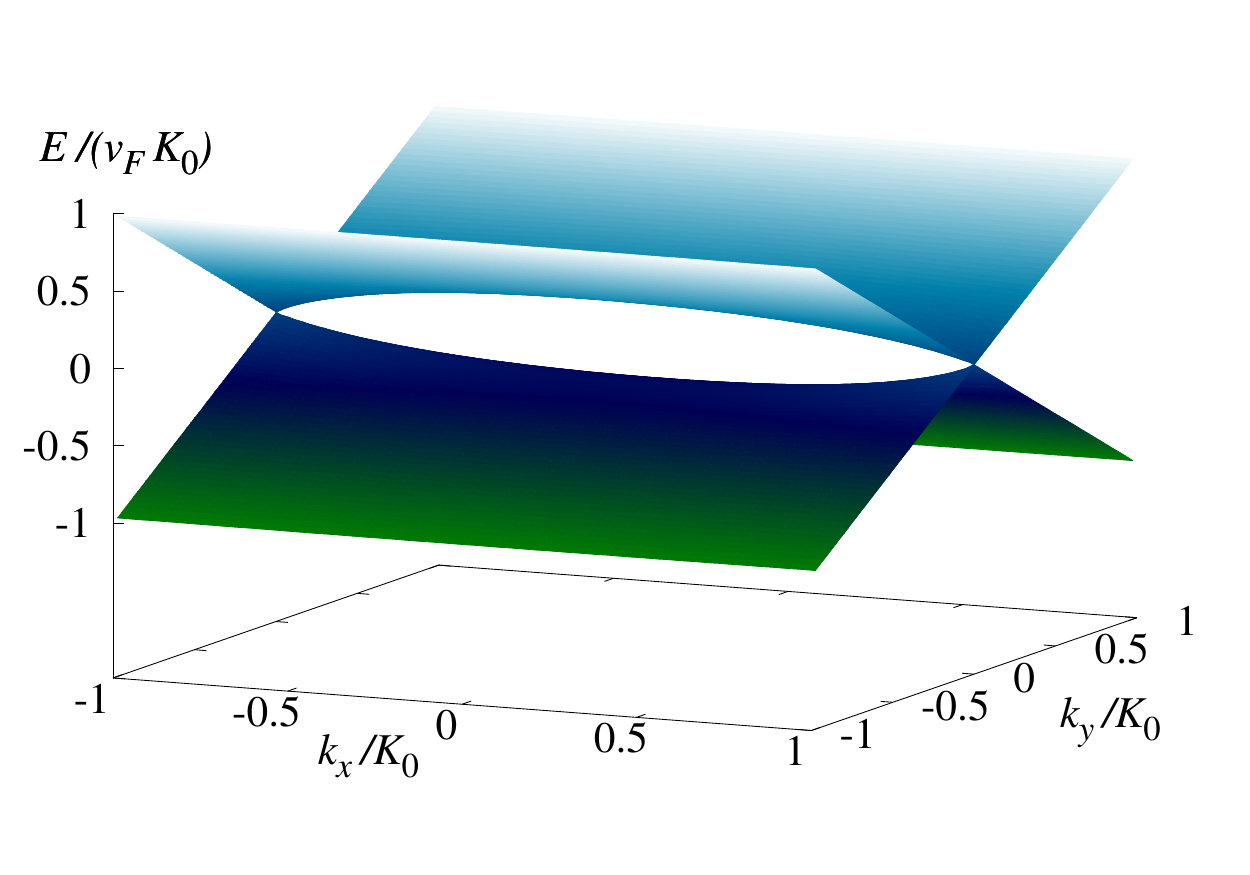}
\caption{
Energy dispersion of surface states of a WSM bilayer in the regime of nonzero
particle-hole coherence. 
Results have been obtained for $T=0$ in the limit $K_0 \ll k_{\text{cut}}$ and $K_0t\ll 1$, 
with $\epsilon_r=10$, $K_0=0.01\,\mathrm{nm}^{-1}$ and $v_F= 100\,\mathrm{meV\,nm}$.}
\label{fig:qp}
\end{figure}

The gap equation (\ref{eq:Delta}) is solved numerically by self-consistent iteration of the MF parameters 
$\Delta_{\bf k}$, which are calculated on a grid of $N_x\times N_y$ points in the reciprocal space
$k_x\in(-K_0,K_0)$, $k_y\in[-k_{\text{cut}},k_{\text{cut}}]$ and are linearly interpolated in between.
We set $t=10$~nm and $k_{\text{cut}}=0.5$~nm$^{-1}$ while keeping $K_0\le0.05$~nm$^{-1}$
so that the conditions $K_0 \ll k_{\text{cut}}$ and $K_0t\ll 1$ are met.
The typical dependence of the gap parameter on the wave vector is shown in Fig.~\ref{fig:D_kxky}.
Consistently with the form of the Coulomb matrix element, the gap parameter
$\Delta_{{\bf k}=k_x \hat{\mathbf{x}}}$ as a function of $k_x$ 
[Fig.~\ref{fig:D_kxky}(a)] is characterized by a maximum for $k_x=0$, 
monotonously decreases with $|k_x|$, and vanishes for $k_x\rightarrow \pm K_0$. 
As a function of $k_y$, the gap parameter shows a peak of width comparable with $K_0$ around $k_y=0$
and decays to zero for larger $|k_y|$, i.e., where the surface-state energy is large compared to $v_FK_0$ [Fig.~\ref{fig:D_kxky}(b)]. 
As shown in Fig.~\ref{fig:qp}, the finite value of the gap parameter renormalizes
the dispersion of the surface states with the opening of an excitonic energy gap for $|k_x|<K_0$. 

\begin{figure}[tb]
\centering
\includegraphics[width=8.6cm]{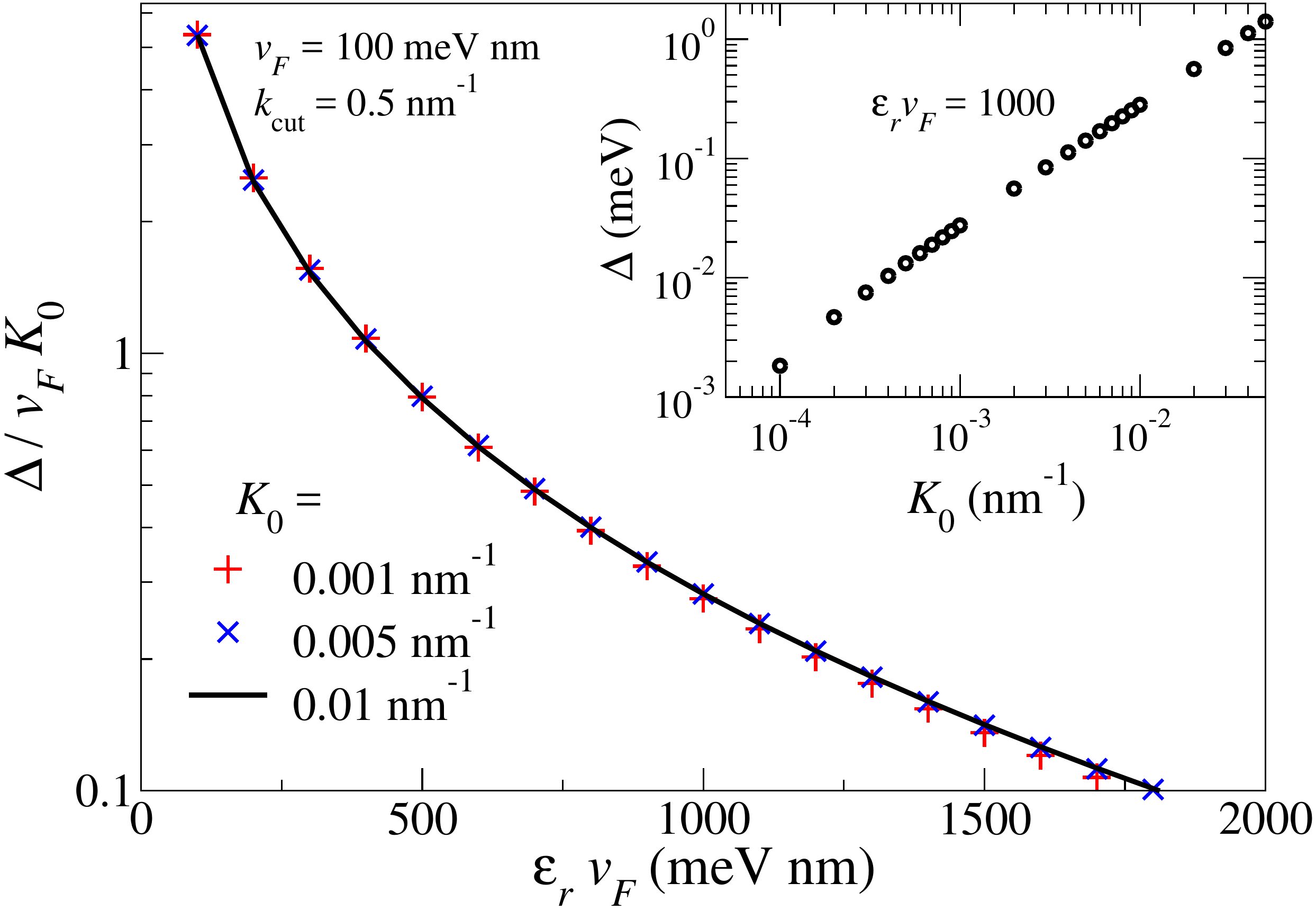}
\caption{Maximum gap parameter $\Delta$ at $T=0$ as a function of
$\epsilon_r v_F$, calculated for various values of $K_0$.  
In the inset, $\Delta$ is plotted as a function of $K_0$ for fixed $\epsilon_r v_F=1000\, \mathrm{meV}\,\mathrm{nm}$.}
\label{fig:D0_er}
\end{figure}

Due to the simple structure of $\Delta_{\bf k}$ evidenced in Fig.~\ref{fig:D_kxky},
it is sufficient to analyze the dependence on system parameters of its
maximum value at ${\bf k}=0$, which we denote by $\Delta$.
If the conditions (\ref{eq:limit}) hold, and taking into account the approximately linear scaling of $\Delta$ with $K_0$, 
we are only left with the temperature $T$ and the effective dielectric constant $\epsilon_r$ as model parameters.
In Fig.~\ref{fig:D0_er}, we study the dependence on $\epsilon_r$
for $K_0=0.001\,\mathrm{nm}^{-1}$, $0.005\,\mathrm{nm}^{-1}$, and $0.01\,\mathrm{nm}^{-1}$ at $T=0$.
As expected, the pairing is favored by small values of the dielectric constant.
The proportionality $\Delta \propto K_0$ is demonstrated by the the collapse of curves calculated 
for different values of $K_0$ over the whole range $\epsilon_r\in[1,20]$
and is further analyzed in the inset, where $\Delta$ is directly plotted as a function of $K_0$ at fixed 
$\epsilon_r v_F=1000\, \mathrm{meV}\,\mathrm{nm}$.

\begin{figure}[tb]
\centering
\includegraphics[width=8.6cm]{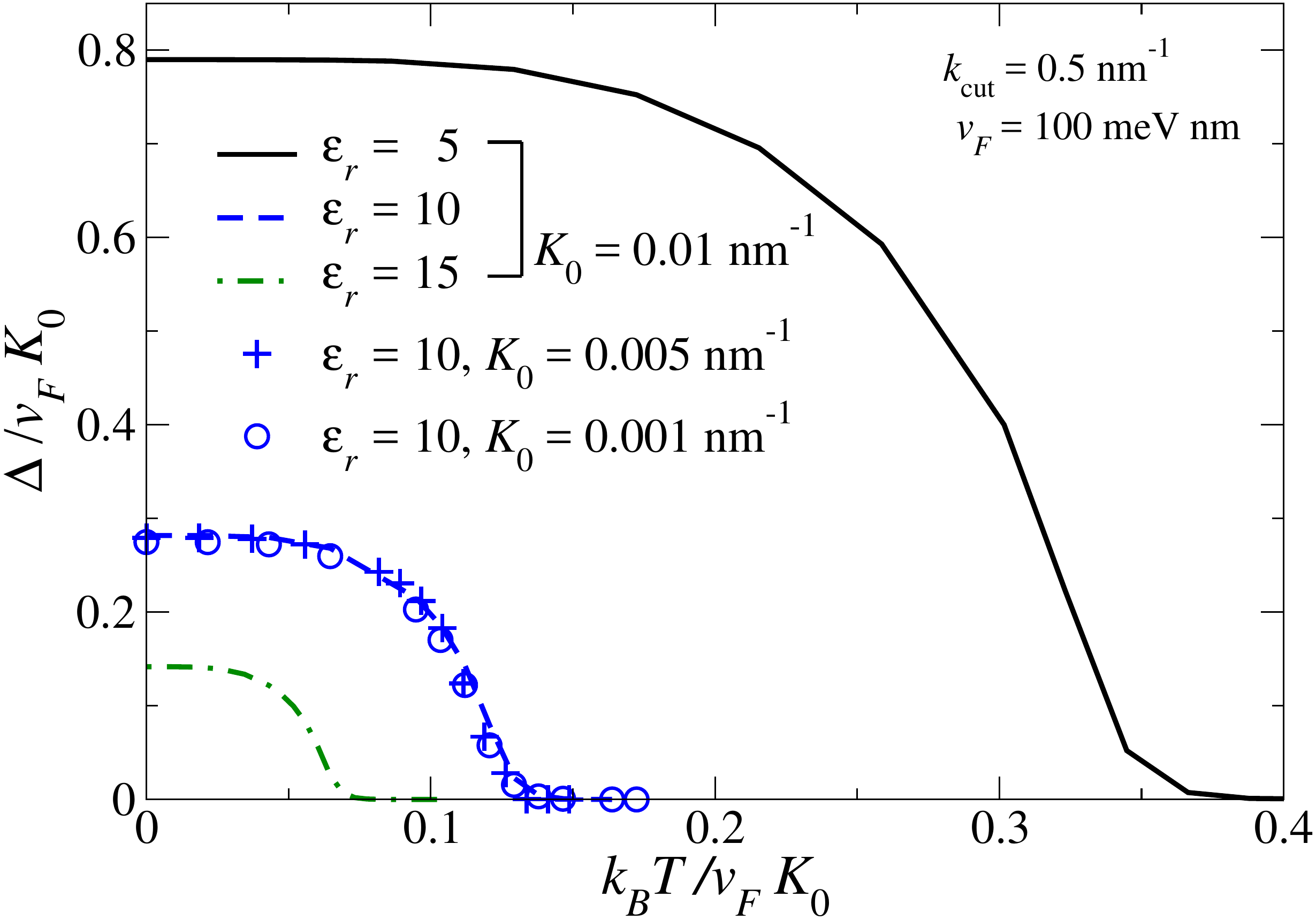}
\caption{Maximum gap parameter $\Delta$ as a function of temperature,
calculated for various values of $K_0$ and $\epsilon_r$.}
\label{fig:D0_T}
\end{figure}

In Fig.~\ref{fig:D0_T}, we analyze the temperature dependence of the
gap parameter $\Delta$.
Curves corresponding to $\epsilon_r=5$, $10$, and $15$ are characterized by
a qualitatively similar dependence on $T$.
However, they shrink towards zero for increasing $\epsilon_r$.
This is indeed compatible with Eq.~(\ref{eq:scaling}), which imposes the
proportionality $\Delta_0 \propto T_c \propto K_0$.
The scaling relation in Eq.~(\ref{eq:scaling}) is further proved by the collapse
of three curves corresponding to different value of $K_0$ at fixed $\epsilon_r=10$. 

\begin{figure}[tb]
\centering
\includegraphics[width=8.6cm]{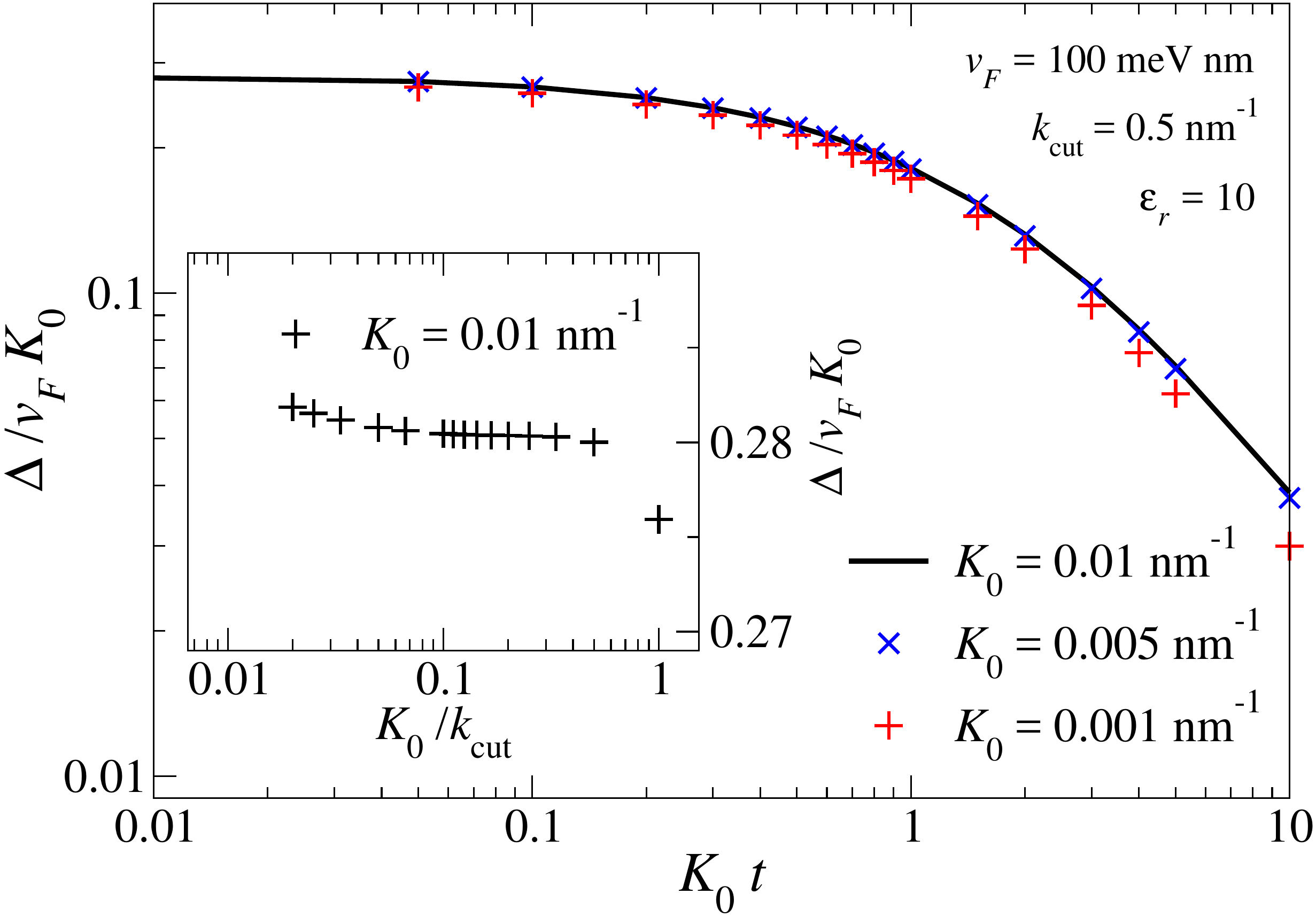}
\caption{Maximum gap parameter $\Delta$ calculated at
$T=0$ as a function of $K_0t$ for three values of $K_0$. 
Inset: $\Delta$ at $T=0$ as a function of $K_0/k_{\text{cut}}$.}
\label{fig:D0_t_K}
\end{figure}

In Fig.~\ref{fig:D0_t_K}, we exhibit the effect of nonzero
$tK_0$ and $K_0/k_{\text{cut}}$ values, which in different ways 
cause a reduction of the effective Coulomb interaction between the layers.
The first parameter leads to an exponential decay of the interaction
on wave-vector scales of $1/t$. 
Typical inter-layer distances could be of the order of $10$--$100\,\mathrm{nm}$
so that this parameter could come to play a significant role for WSMs with 
large enough $K_0$ values.
The second parameter $K_0/k_{\text{cut}}$ depends on $k_{\text{cut}}$, which represents
the limit of the allowed values of $|k_y|$ and $|k_y'|$ in the gap equation
(\ref{eq:Delta}). 
Figure \ref{fig:D0_t_K} shows that this parameter is less effective in reducing the interaction.

\begin{figure}[tb]
\centering
\includegraphics[width=8cm]{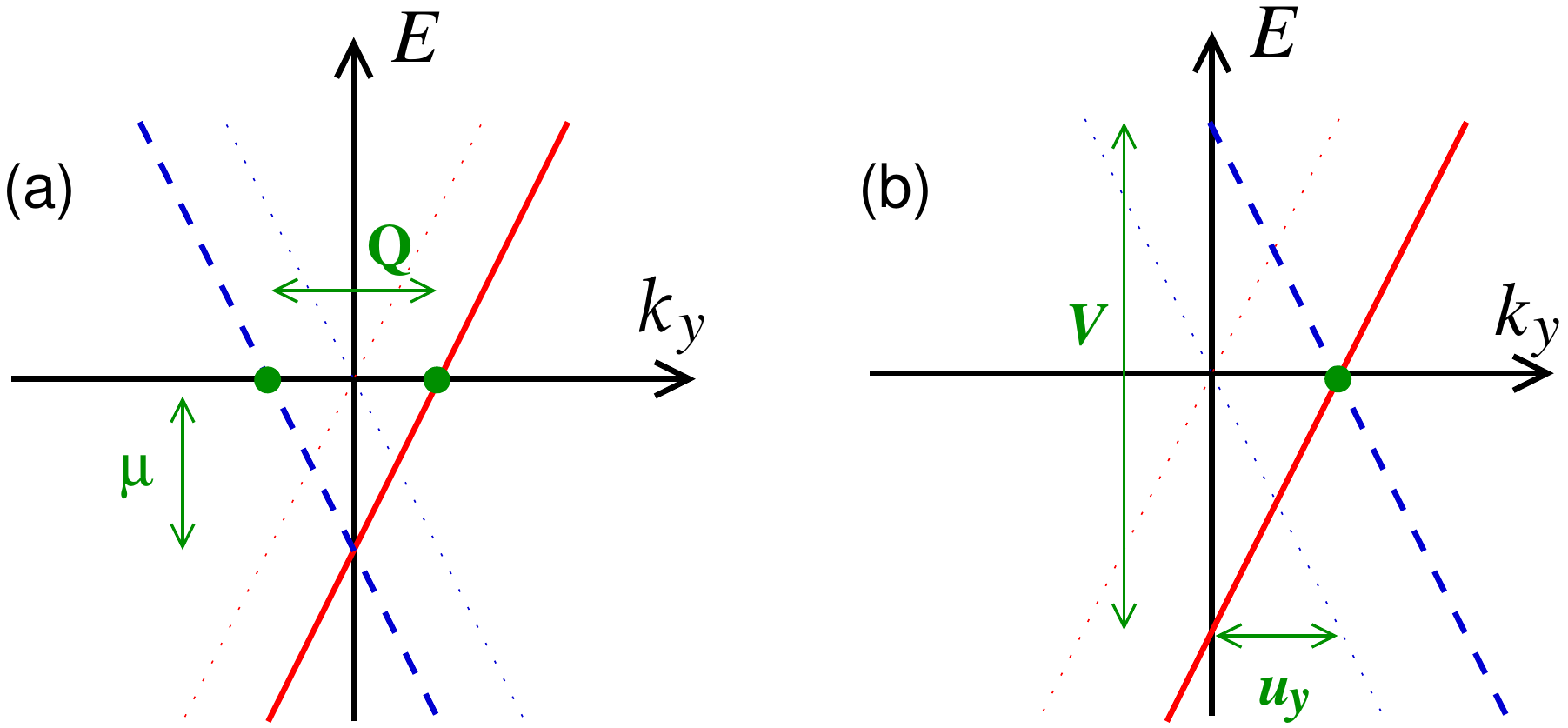}\\
\includegraphics[width=5cm]{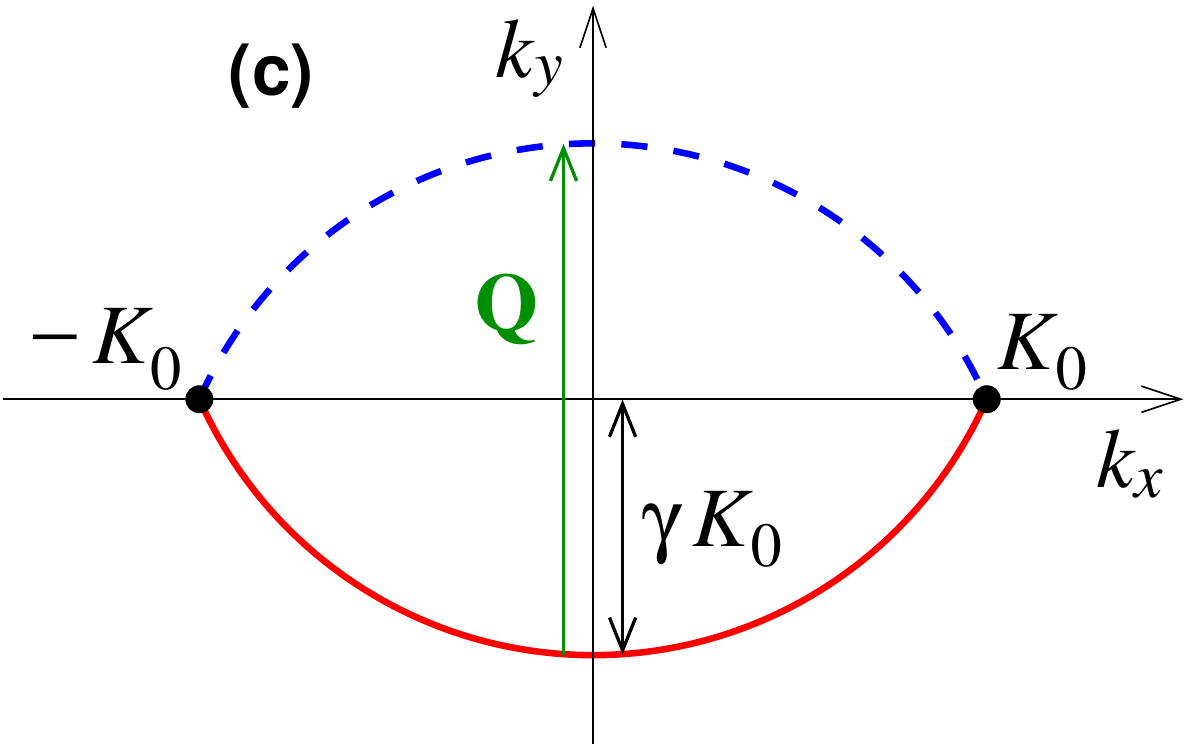}
\caption{Dispersion curves of the FASs for (a) nonzero chemical potential
$\mu$ and (b) nonzero inter-layer potential bias $V$.
(c) Model of curved FAs. In all panels, dashed and full lines refer to surface states belonging to opposite surfaces.}
\label{fig:mu_V}
\end{figure}

Above, we have presented the results obtained at chemical potential $\mu=0$.
Now, we consider the case of $\mu\neq 0$, where the WSM contains small Fermi pockets
in the bulk. 
We assume that $|\mu|$ is sufficiently small to justify neglecting the finite
density of states of bulk states in the description of electron-hole pairing.
As shown in Fig.~\ref{fig:mu_V}(a), a finite $\mu$ leads to the introduction
of a nesting vector ${\bf Q}=\mu/v_F \hat{\mathbf{y}}$.
Since the Coulomb matrix element in Eq.~(\ref{eq:V_Q=0}) is independent of
$Q_y$, the analysis remains essentially unchanged.
[Minor changes are due to the modified integration domain in the gap equation
(\ref{eq:Delta}) and do not play any role as long as
$|\mu|/v_F\ll K_0,k_{\text{cut}}$.]
The result is a gap parameter $\Delta_{\mathbf{k},\mathbf{Q}}$ and thus a pair amplitude
$F_{\mathbf{k},\mathbf{Q}}$ with a nonzero wave vector ${\bf Q}=\mu/v_F \hat{\mathbf{y}}$, 
while the gap amplitude $\Delta$ has the same dependence on parameters as for $\mu=0$.

The effect of an applied potential energy bias $V$ between the two WSM layers (at $\mu=0$) is described in Fig.~\ref{fig:mu_V}(b) 
and consists in a vertical displacement between the energy dispersions of the surface states in the \textit{A} and \textit{B} layers.
Due to the linearity of the dispersion, this is equivalent to a
shift of the FAs in the \emph{same} direction by the wave vector
$u_y \hat{\mathbf{y}}$ with $u_y = V/2v_F$.
The nesting vector is thus $\mathbf{Q}=0$.
As long as $|u_y|\ll k_{\text{cut}}$, the MF solution of the electron-hole
pairing problem leads to the previously discussed results for $\Delta_{{\bf k},{\bf Q}=0}$ with the 
caveat that $k_y$ has to be replaced by $k_y-u_y$.

\begin{figure}[tb]
\centering
\includegraphics[width=8.6cm]{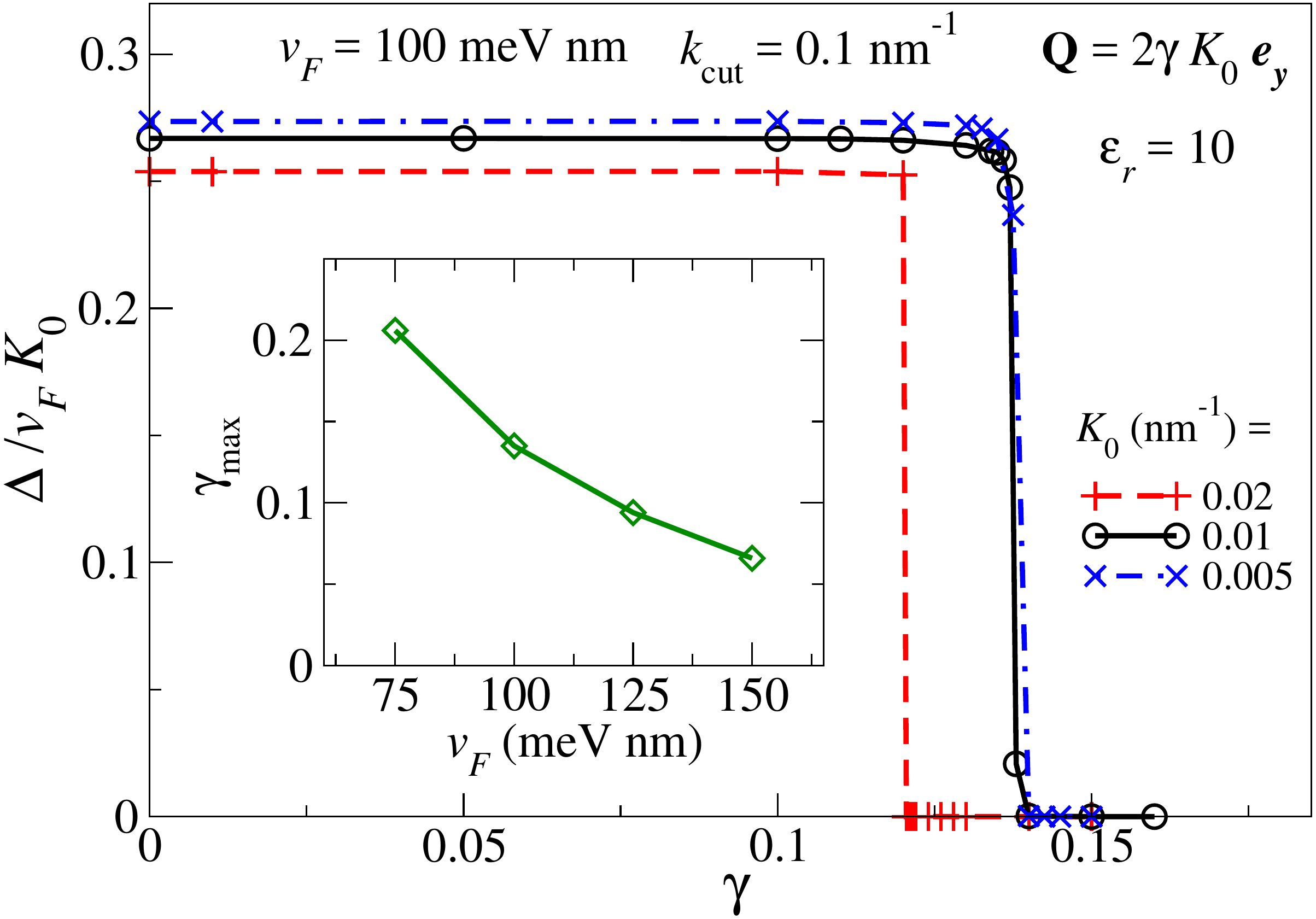}
\caption{Maximum gap parameter $\Delta$
at $T=0$ for a model with curved FAs as a function of the curvature parameter $\gamma$ 
for various values of $K_0$, tuning $\bf Q$ to the optimal nesting conditions.
Inset: maximal value of the curvature parameter compatible with electron-hole coherence as a function of the Fermi velocity $v_F$.}
\label{fig:gamma}
\end{figure}

The WSM model introduced in Sec.~\ref{sec:WSM} possesses particle-hole symmetry.
Such a symmetry would be accidental in real WSMs.
For example, the allowed additional terms
$H''=B k^2 \, \sigma_0\otimes\tau_x-D\, \sigma_0\otimes\tau_0$, with $|D|<|B|$, 
introduce quadratic corrections to the model, bending the bands at high
energies and breaking particle-hole symmetry for $D\ne 0$.
The numerical solution for the surface states in a slab geometry shows that the FA
is curved for $D\ne 0$, while the dispersion curve in the direction perpendicular
to the FA remains approximately linear. 
In order to quantify the effect of the FA curvature in a transparent and tractable case, 
we consider a FA with constant curvature, i.e., a circular arc with dispersion
\begin{equation}
 \varepsilon^{(\pm)}_{\bf k} 
  = \pm v_F\, \big[\big| {\bf k} \mp (R-\gamma K_0) \hat{\mathbf{y}}\big|-R\big]
\label{eq:dispC}
\end{equation}
for $k_x\in(-K_0,K_0)$, where $R=(\gamma^2+1) K_0/(2\gamma)$
is the radius of curvature in terms of the curvature parameter
$\gamma\ge 0$ and the FAs on opposite surfaces are distinguished by the sign $\pm$.
The curved FA is shown in Fig.~\ref{fig:mu_V}(c).
Clearly, for $\gamma=0$ this dispersion relation reproduces the
straight FA of our original model in Eq.~(\ref{eq:dispFASS}).
The curvature evidently reduces the nesting. 
The optimal nesting vector is given by ${\bf Q}=2\gamma K_0 \hat{\mathbf{y}}$ 
and it is natural to consider a pair amplitude $F_{\mathbf{k},\mathbf{Q}}$
with this modulation vector.
The results obtained by solving the gap corresponding equation are summarized in Fig.~\ref{fig:gamma}. 
We find that electron-hole pairing suddenly disappears for a curvature parameter 
exceeding a maximal value $\gamma_{\text{max}}$ on the order of $1.4$ for $v_F=100\,\mathrm{meV\,nm}^{-1}$,
weakly increasing for decreasing $K_0$ in the range $K_0\in[0.005\,\mathrm{nm}^{-1},0.02\,\mathrm{nm}^{-1}]$. 
In the inset, we show the dependence of $\gamma_{\text{max}}$ on the Fermi velocity $v_F$.
Lower $v_F$ alleviate the effect of the FA curvature, increasing the maximum value of the curvature compatible with electron-hole coherence.  
The sudden transition as a function of the curvature parameter $\gamma$ suggests it to be of first order.
We have checked that at the transition point, both the condensate phase and the $\Delta=0$ free phase are stable 
MF solutions with the same free energy, which shows that the transition is indeed of first order.

An intuitive explanation for the sudden drop of the coherence above
$\gamma_{\text{max}}$ can be based on the curved FAs being not perfectly nested 
for $k_x\ne0$: the energy displacement of surface states separated by $\mathbf{Q}$
is proportional to $\gamma v_F k_x$, which is an energy scale competing with
$\Delta$ in the gap equation (\ref{eq:Delta}).
On the other hand, our previous analysis suggests that 
$\Delta\propto \tilde{K}_0$, where $\tilde{K}_0<K_0$ describes the
effective extent of the coherence on the FA. 
If the energy displacement prevails at large $k_x$, the region with
sizable gap $\Delta$ on the FA shrinks (smaller $\tilde{K}_0$),
which leads to a decrease of the self-consistent value
of $\Delta$ itself. 
Hence, the energy displacement will also prevail at smaller $k_x$, leading to
a decrease of $\tilde{K}_0$ and further reduction of $\Delta$.
This causes a positive feedback, which wipes out coherence on the whole FA.
From this analysis for a simplified model for curved FAs with constant curvature, we infer that particle-hole pairing is favored
in WSM bilayers between relatively straight ($\gamma<\gamma_{\text{max}}$) portions of FAs, 
which are well nested with an arbitrary nesting vector ${\bf Q}$.

\begin{figure}[tb]
\centering
\includegraphics[width=8.0cm]{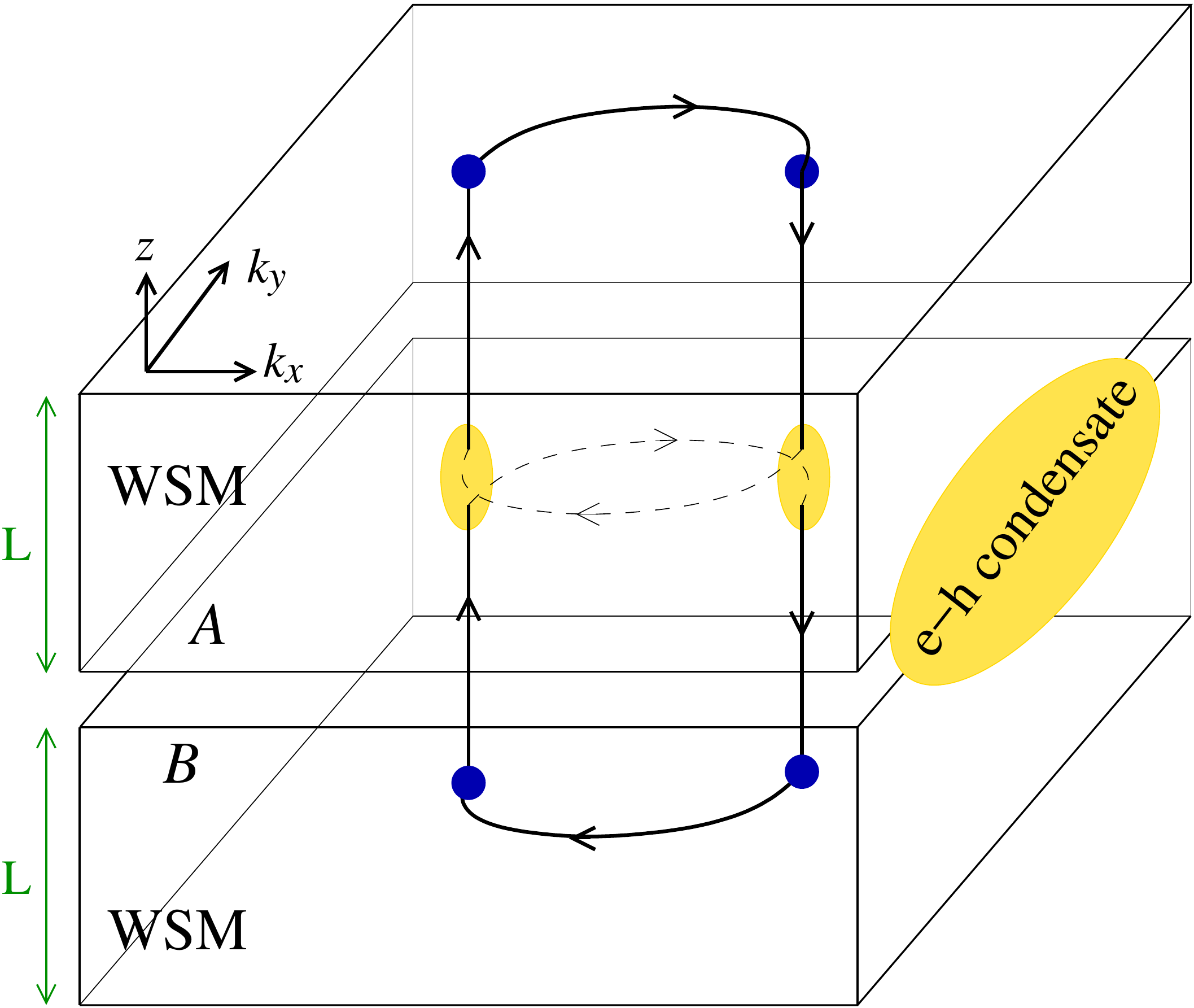}
\caption{Sketch of a magnetic orbit extending over the two WSM
layers connected by the electron-hole condensate.
The orbit includes FAs at the outermost surfaces and vertical bulk
portions, see the text and Ref.~\cite{potter}.}
\label{fig:QO}
\end{figure}

Finally, it is of course important to identify experimental signatures
of electron-hole pairing in FAs of a WSM bilayer structure.
Potter \textit{et al.}~\cite{potter} have proposed that the FAs in WSM
realize an anomalous closed magnetic orbit, where the opposite surfaces are connected 
through the bulk of the WSM.
These orbits can lead to observable quantum oscillations as a function of 
the inverse magnetic induction, $1/B$, of frequency $\Omega_{B^{-1}}\approx e\pi v_F/\mu K_0$
in several physical quantities, such as the conductivity and the
magnetization.
These oscillations are expected to persist down to $1/B_{\rm sat}$, where $B_{\rm sat}\approx K_0/L$ and $L$ is the 
thickness of the sample~\cite{potter}.

How does this picture change for our bilayer system with electron-hole
condensation at the interface?
In Fig.~\ref{fig:QO}, we sketch the modified orbits in the WSM bilayer.
The electron-hole condensate gaps out the FAs at the facing WSM
surfaces so that the electrons cannot complete their orbits along these arcs.
On the other hand, the condensate can absorb an electron in layer \textit{A}
and emit an electron in layer \textit{B} (or vice versa). 
This is the excitonic analogue of crossed Andreev reflection,
where an electron in the opposite layer takes
the place of a hole with opposite spin. 
Hence, the condensate connects the bulk portions of the anomalous magnetic orbits,
leading to closed orbits extending over the whole bilayer structure.  
As a consequence, the effective thickness of the sample is $2L$.
In the absence of the condensate, the anomalous orbits are closed in each layer separately and the thickness is $L$. 
Hence, $B_\mathrm{sat}$ is halved when the condensate is present.
The disappearance of quantum oscillations in the field range between $K_0/2L$ and $K_0/L$ can
therefore serve as a signature of the coherence between the two surface FAs.

\section{Summary and conclusions}  
\label{sec:conclusion}

In conclusion, we have studied electron-hole pairing between facing
FASs of a WSM bilayer.
We propose that a coherent electron-hole condensate can be realized
between relatively straight portions of the FAs.
The condensate shows a modulation in space with the modulation vector
given by the optimal nesting vector $\mathbf{Q}$ between the FAs.
While WSMs with a single FA have recently been predicted \cite{wang2016},
most systems feature multiple pairs of Weyl nodes. 
The scenario then becomes more complex with the possible pairing of relatively straight
portions of the FAs belonging to all different pairs of Weyl points nested by
different wave vectors ${\bf Q}_1$, ${\bf Q}_2$, etc., leading to the coexistence
of several nonzero pair amplitudes $F_{\mathbf{k},\mathbf{Q}_1}$,
$F_{\mathbf{k},\mathbf{Q}_2}$, etc.
The condensation leads to the gapping of the surface states.
We find that the gap and the critical temperature are proportional
to the effective extension of the straight FA intervals. 
The presence of the condensate could be detected from its effect on
the anomalous quantum oscillations peculiar to Weyl semimetals~\cite{potter}.
Essentially, it causes the full bilayer to behave like a single WSM layer
and this effect could be switched on and off by tuning through the
condensation transition.
In addition, the condensation could in principle 
lead to superfluid behavior of the electron-hole pairs (dipolar superfluid),
where counter-propagating electron and hole 
supercurrents could be generated by suitable electric~\cite{su} or
magnetic~\cite{balatsky} fields.

\acknowledgments

We thank C. Berke, M. Vojta and B. Trauzettel for helpful discussions.
Financial support by the Deutsche Forschungsgemeinschaft through Collaborative Research Center SFB 1143 is gratefully acknowledged.

\appendix

\onecolumngrid

\section{Quasi-2D Coulomb interaction between FASs}
\label{app:V}

In this appendix, we derive the Coulomb-interaction matrix elements between FASs.
For our layered model, it is useful to Fourier transform the Coulomb potential in the
directions parallel to the interfaces but leave it in real space for the perpendicular
\textit{z}-direction \cite{haug}, 
\begin{equation}
 V_{\bf q}(z) = \frac{e^2}{2 \epsilon_0\epsilon_r L^2}\, \frac{1}{q}\, e^{-q |z|},
\label{eq:V2D}
\end{equation}
where ${\bf q}$ is the in-plane wave vector.  
The partially Fourier-transformed charge density can be obtained from
the FAS wave functions of Eq.~(\ref{eq:FA_wavef}), 
\begin{equation}
\rho_{{\bf k} + {\bf q},{\bf k}}^{(\pm)}(z) = \int dx\int dy\,
    e^{i {\bf q}\cdot {\bf r}} \,
    \Psi_{{\bf k} + {\bf q}}^{(\pm)*}({\bf r})\,
    \Psi_{\bf k}^{(\pm)}({\bf r}) 
  = \left[\frac{K_0^2 - (k_x+q_x)^2}{K_0}\,
    \frac{K_0^2-k_x^2}{K_0}\right]^{1/2}
  e^{\pm 2K_0z}\, \cosh{[(2k_x\hspace{-0.1cm}+\hspace{-0.1cm}q_x)z]}\,
  \Theta(\mp z),
\label{eq:CD} 
\end{equation}
where the sign $\pm$ refers to the surface with normal direction $\pm \hat{\mathbf{z}}$,
i.e., to the surface \textit{A} and \textit{B}, respectively,
and $\Theta$ is the Heaviside step function.
Using Eqs.~(\ref{eq:V2D}) and (\ref{eq:CD}), we can now compute the Coulomb
matrix element between FASs on opposite surfaces \textit{A} and \textit{B}
separated by a spacer of thickness $t$. 
Due to the presence of the spacer, the charge density for surface \textit{B} ($-$) is shifted by $t$ in $z$-direction. 
The matrix element reads as
\begin{align}
 V_{\mathbf{q},\mathbf{k},\mathbf{k}'} &=  \int dz \int dz'\,
   V_{\bf q}(z-z')\, \rho_{{\bf k} +{\bf q},{\bf k}}^{(+)}(z)\,  
      \rho_{{\bf k'} -{\bf q},{\bf k'}}^{(-)}(z'-t) \nonumber\\
   &= \frac{e^2}{2 {\epsilon_0\epsilon_r}}\, \frac{e^{-qt}}{q}\,
   \sqrt{\frac{[K_0^2-(k_x+q_x)^2](K_0^2-k_x^2)[K_0^2-(k_x'-q_x)^2](K_0^2-k_x'^2)}
     {[(q+2K_0)^2-(2k_x+q_x)^2]^2 [(q+2K_0)^2-(2k_x'-q_x)^2]^2}}\:
   \frac{(2q+4K_0)(2q+4K_0)}{K_0^2} .
\end{align}
The matrix element appearing in Eq.~(\ref{eq:H_C_pair}) is now
\begin{equation}
 V_{ {\bf k} , {\bf \tilde{k}} }^{\bf Q} =
  V_{\mathbf{q} = \tilde{\mathbf{k}}-{\bf k},\mathbf{k},\mathbf{k}'=\tilde{\mathbf{k}}+{\bf Q}}, 
\end{equation}
which for the special case $Q_x=0$ simplifies to Eq.~(\ref{eq:V_Q=0}).

\section{Scaling form of the gap equation}
\label{app:I}

We assume here the gap function $\Delta$ to be constant along the FA, 
for the case of straight FAs displaced with respect to each other
by ${\bf Q}=\mu/v_F \hat{\mathbf{y}}$.
The quasiparticle energies in Eq.~(\ref{eq:disp_qp}) are given by
\begin{equation}
 E^\pm_{\bf k} \,=\, \pm \sqrt{v_F^2 k_y^2+ \Delta^2}.
\end{equation}
We calculate $\Delta$ at ${\bf k}=0$ using the gap equation (\ref{eq:Delta}), where we pass
from a sum over ${\bf k'}$ to an integral over reciprocal space,
\begin{align}
  1 &= \frac 12 \int_{-K_0}^{K_0} \!dk_x \int_{-k_{\text{cut}}}^{k_{\text{cut}}}\!dk_y \,V_{0,\bf k}^{Q_x=0}\,
                                         \frac{\tanh{\frac{\sqrt{v_F^2 k_y^2+ \Delta^2}}{2k_B T}}}{\sqrt{v_F^2 k_y^2 +\Delta^2}}, 
\end{align}
where $k_{\text{cut}}$ parametrizes the extension of the uniaxial linear
dispersion along $k_y$ of the FASs and we have taken advantage of the
identity $f(z)-f(-z)=-\tanh{(z/2)}$  for the Fermi function.
Because of symmetry, we can restrict the integration in the region
$k_x\in[0,K_0]$ and $k_y\in[0,k_{\text{cut}}]$ and multiply the result by $4$.
We now change the integration variables to the dimensionless
$x=k_x/K_0$ and $y=k_y/k_{\text{cut}}$, introducing also $r=\sqrt{x^2+y^2}$, 
and arrive at the final expression for Eq.~(\ref{eq:Delta}),
\begin{align}
 \frac{\pi^2 \epsilon_0\epsilon_r v_F}{e^2}
   &= \int^1_0 \!dx \int^{k_{\text{cut}}/K_0}_0\! dy\:
      \frac{e^{-r K_0 t}}{r}\,
      \frac{(r+2)^2 (1-x^2)}{[(r+2)^2-x^2]^2}\,
      \frac{\tanh \frac{\sqrt{y^2+(\Delta/v_FK_0)^2}}{2\,(k_BT/v_FK_0)}}
      {\sqrt{y^2+(\Delta/v_FK_0)^2}} \nonumber \\
&= I\left( \frac{\Delta}{v_FK_0}, \frac{k_BT}{v_FK_0};
  \frac{k_{\text{cut}}}{K_0}, \frac{1}{K_0 t} \right).
\end{align}
Note that the integral is well defined also in the limit $k_{\text{cut}}\to\infty$.



\twocolumngrid

\end{document}